\documentclass[aps,pre,twocolumn]{revtex4}
\usepackage{amsmath}
\usepackage{amssymb}
\usepackage{graphicx}
\usepackage{hyperref}
\usepackage{color}
\usepackage{physics}
\usepackage[caption=false]{subfig}
\usepackage[capitalize]{cleveref}

\newcommand{\N}[1]{\mathcal{N}_{#1}}
\newcommand{\Pins}[1]{P_{\mathrm{ins},#1}}
\newcommand{\pore}[1]{\Phi_{#1}}
\newcommand{\distance}[1]{D_{#1}}
\newcommand{\surface}[1]{S_{#1}}
\newcommand{\volume}[1]{V_{#1}}
\newcommand{\vtot}{V_\mathrm{tot}}
\newcommand{\unit}{V_d}
\newcommand{\myset}{\mathbb{S}}
\newcommand{\fractal}{\gamma}
\newcommand{\binomial}[2]{\mathrm{C}^{#2}_{#1}}

\graphicspath{{figures/}}

\DeclareMathOperator{\erfc}{Erfc}

\begin{document}

\title{Mean-field approach to Random Apollonian Packing}

\author{Pierre Auclair} \email{pierre.auclair@uclouvain.be}
\affiliation{Cosmology, Universe and Relativity at Louvain (CURL),
  Institute of Mathematics and Physics, University of Louvain, 2 Chemin
  du Cyclotron, 1348 Louvain-la-Neuve, Belgium}

\date{\today}

\begin{abstract}
    We revisit the scaling properties of growing spheres randomly seeded in $d=2,3$ and $4$ dimensions using a mean-field approach.
    We model the insertion probability without assuming \emph{a priori} a functional form for the radius distribution.
    The functional form of the insertion probability shows an unprecedented agreement with numerical simulations in $d=2, 3$ and $4$ dimensions.
    We infer from the insertion probability the scaling behavior of the Random Apollonian Packing and its fractal dimensions.
    The validity of our model is assessed with sets of $256$ simulations each containing 20 million spheres in $2, 3$ and $4$ dimensions.
\end{abstract}

\maketitle

\section{Introduction}

Bubble nucleation is a phenomenon ubiquitous in physics, with applications ranging from the geometry of tree crowns \cite{horn1971adaptive}, the structure of porous media \cite{van1996network} and the generalized problem of sphere packing \cite{conway2013sphere} to the description of cosmic voids \cite{Gaite:2002mq, Gaite:2005di, Gaite:2006tr} and the signatures of cosmological phase transitions in terms of topological defects \cite{Borrill:1995gu} and in gravitational waves \cite{Kosowsky:1991ua, Auclair:2022jod}.

In this article, we study the universal properties of a simple, yet broad class of sphere packing models dubbed ``Packing-Limited-Growth'' (PLG).
Such mechanisms entail objects being seeded randomly, growing and stopping upon collision with other objects.
A simple model of PLG is the ABK model, named after Andrienko, Brilliantov and Krapivsky \cite{andrienko1994pattern,brilliantov1994random}.
In their setting, $d$-dimensional spheres are seeded randomly in space and time and grow linearly with time.
They determine the fractal dimension for $d = 1$ and make a prediction for higher dimensions
\begin{equation}
    \fractal \approx d + 1-d \exp[\frac{2 \qty(d-2^{d+1}+3)}{d+2}] =
    \begin{cases}
        2.554 & d=2 \\
        3.945 & d=3 \\
        4.999 & d=4
    \end{cases},
\end{equation}
independently of the growth velocity.
More generally, it is claimed that the fractal dimension is independent of the specifics of the growth dynamics \cite{PhysRevE.65.056108} and the shape of the objects \cite{delaney2008relation}.

\begin{figure}
    \includegraphics[width=.4\textwidth]{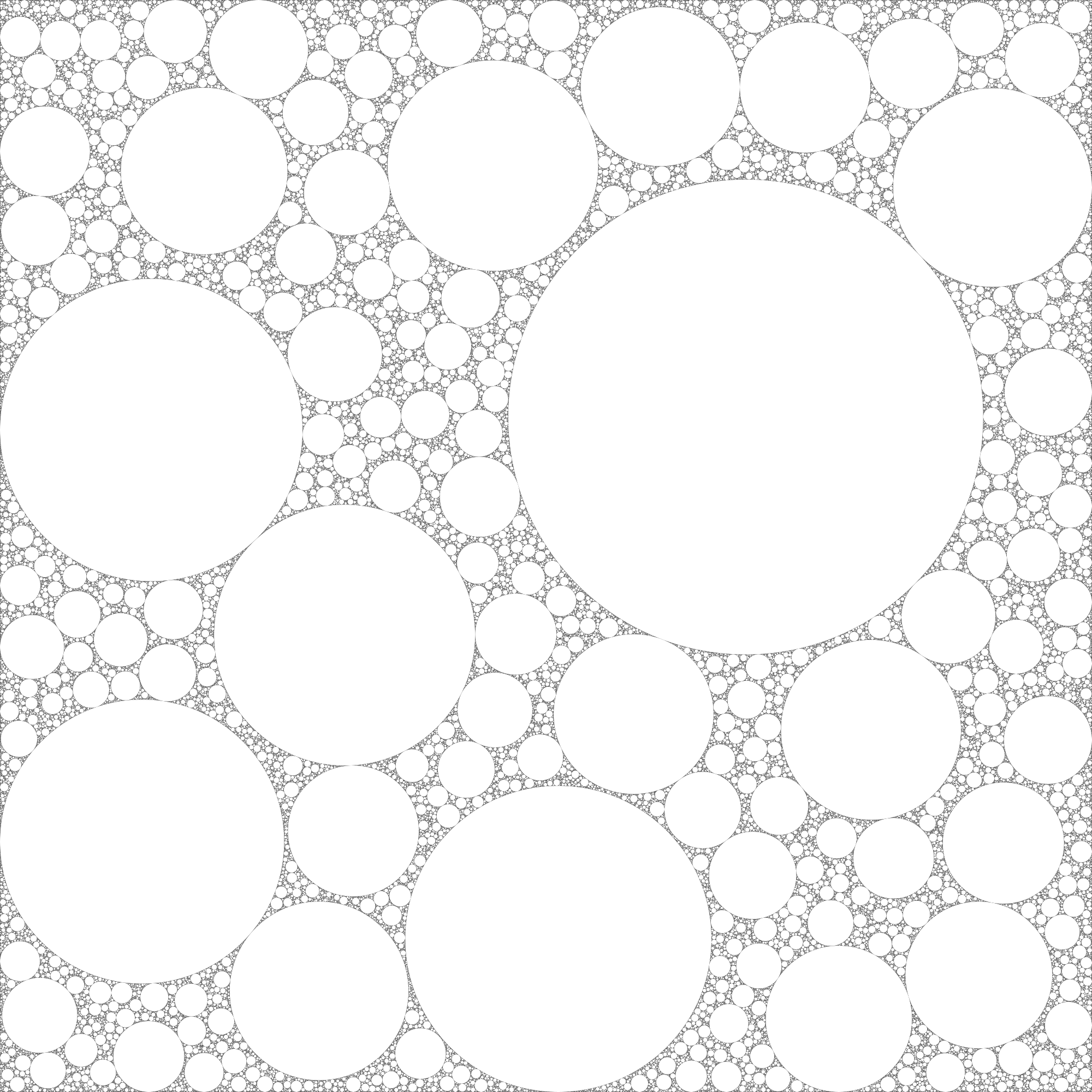}
    \caption{Visualization of a $2d$ RAP containing $10^6$ spheres.}
    \label{fig:sim}
\end{figure}

In this article, we examine a related mechanism referred to as ``Random Apollonian Packing'' (RAP) and illustrated in \cref{fig:sim} \cite{manna1992space,manna1991precise}.
Here, $d$-dimensional spheres are seeded one at a time randomly in space in a finite-sized volume $\vtot$, and take the largest possible radius that avoids overlap.
This mechanism is inspired by the well-known Apollonian packing \cite{mandelbrot1982fractal,amirjanov2012fractal} and is a limit of the ABK model when the growth velocity is infinitely large.

The interest of the RAP mechanism is that it is thought to share universal properties with the more general ABK mechanism but can be approached from a completely different angle.
Namely, the ABK mechanism is dynamical -- multiple spheres are growing at the same time and collide with one another -- whereas the RAP mechanism is sequential -- spheres are added one at a time in a static environment.
In this sense, our work intends to improve on Ref~\cite{PhysRevE.65.056108} in that we also model the insertion probability of spheres.

In \cref{sec:mean-field}, we present our mean-field approach to model the cumulative insertion probability $\Pins{n}(r' > r)$.
We show in \cref{sec:large-n} how to calculate the fractal dimension in this framework.
Then, \cref{sec:ud,sec:it} present two approximations with increasing accuracy for the ``surface model'' and the computation of the fractal dimension.
Finally, we assess the validity of our model with numerical simulations in $d=2, 3$ and $4$ in \cref{sec:numerical} and make the connection with Ref~\cite{PhysRevE.65.056108}.

\section{Mean-field approach}
\label{sec:mean-field}

\begin{figure}
    \includegraphics[width=.4\textwidth]{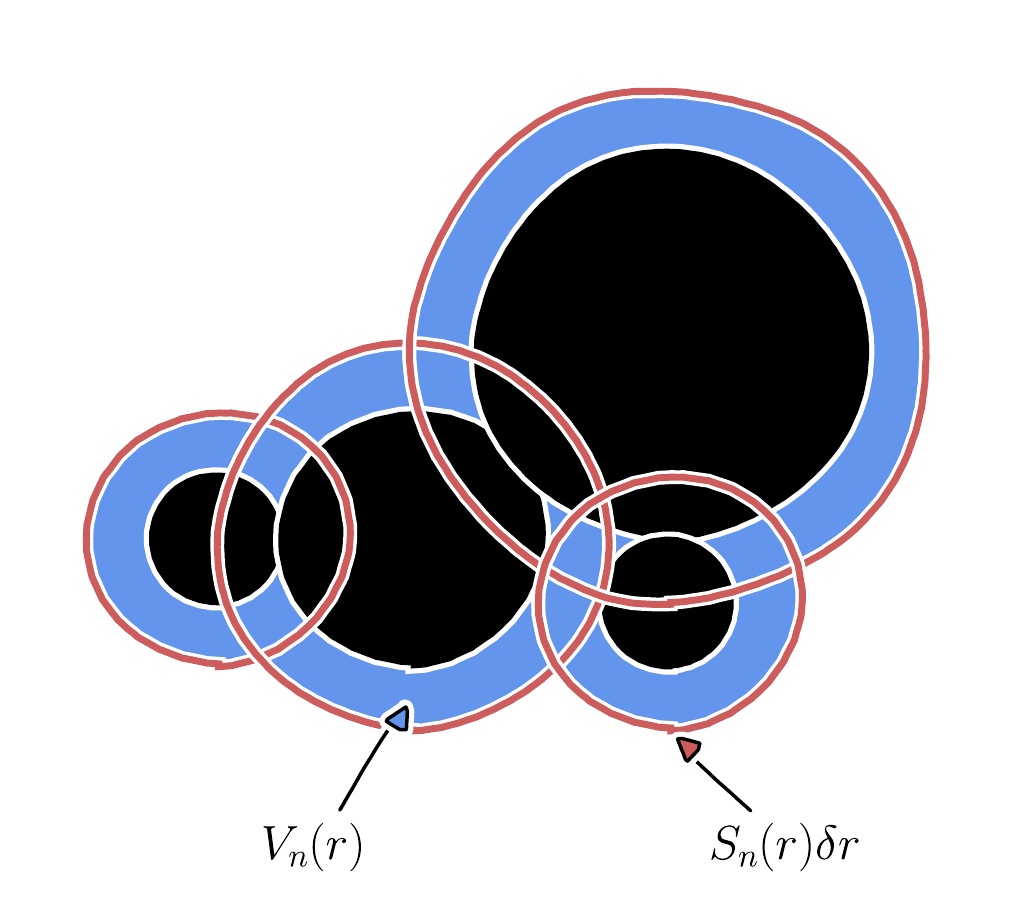}
    \caption{
        The volume $\volume{n}(r)$ (in blue) contains all the points with a distance $\in [0, r]$ from the $n$ preexisting spheres.
        The dependence of this volume with respect to $r$ consist in adding successive layers of thickness $\delta r$ and ``surface'' $\surface{n}(r)$ (in red).
        A fraction of these layers is already included in $\volume{n}(r)$.
        In this article, we estimate that this fraction is $\volume{n}(r) / \pore{n}$, with $\pore{n}$ the remaining pore space.
    }
    \label{fig:picture}
\end{figure}

We start by considering the cumulative insertion distribution function $\Pins{n+1}(r' > r)$, the probability to insert the $n+1$-th sphere with a radius larger than $r$ in $d$ dimensions.
The cumulative insertion distribution can be written as a function of $\pore{n}$, the remaining pore space after the $n$-th sphere insertion, and $\volume{n}(r)$ defined as
\begin{equation}
    \volume{n}(r) = \int_{\pore{n}} \Theta[r - \distance{n}(\vb{x})] \dd[d]{\vb{x}},
\end{equation}
the empty volume located at a distance $\distance{n}(\vb{x})$ less than $r$ from the $n$ first spheres.
The integral is performed over the remaining pore.
The cumulative insertion distribution reads
\begin{equation}
    \Pins{n+1}(r' > r) = 1 - \frac{\volume{n}(r)}{\pore{n}}.
\end{equation}
Indeed, this expression satisfies the properties of RAP
\begin{itemize}
    \item a sphere with radius $r' < r$ can only be inserted in $\volume{n}(r)$, otherwise it would not be tangent to one of the $n$ first inserted spheres and it would continue to grow
    \item reciprocally, a sphere with radius $r' > r$ cannot be inserted in $\volume{n}(r)$, otherwise it would overlap with one of the $n$ first inserted spheres
    \item it naturally vanishes when $\volume{n}(r) = \pore{n}$, that is when there is no more space available at a distance greater than $r$ from the $n$ first inserted spheres
\end{itemize}
As illustrated in \cref{fig:picture}, the geometry of $\volume{n}(r)$ is very complex for $d \geq 2$, and the purpose of this article is to provide an approximation for this function.

As a mean-field approximation we estimate that the volume $\volume{n}(r)$, as a function of $r$, grows by adding successive infinitesimal layers of thickness $\delta r$ and ``surface'' $\surface{n}(r)$.
Each of these new layers has an independent probability to be already included in the volume $\volume{n}(r)$, see \cref{fig:picture}.
We make the assumption that a fraction $\volume{n}(r) / \pore{n}$ of these layers is already accounted for
\begin{equation}
    \volume{n}(r + \delta r) = \volume{n}(r) + \surface{n}(r) \delta r \qty[1 - \frac{\volume{n}(r)}{\pore{n}}].
\end{equation}
In the limit $\delta r \rightarrow 0$, $\Pins{n+1}$ is solution to an ordinary differential equation
\begin{equation}
    \dv{\Pins{n+1}(r'>r)}{r} = - \frac{\surface{n}(r)}{\pore{n}} \Pins{n+1}(r' > r),
    \label{eq:ode}
\end{equation}
which can be integrated
\begin{equation}
    \Pins{n+1}(r'>r) = \exp[- \int_0^r \frac{\surface{n}(r')}{\pore{n}} \dd{r'}].
    \label{eq:master}
\end{equation}
As expected, $\Pins{n+1}(r' > 0) = 1$ and the insertion probability naturally vanishes at large $r$.

In this framework, we define the radius cumulative distribution after the $n$-th insertion $\N{n}(r'>r)$ as the sum of the $n$ first insertion cumulative distributions
\begin{equation}
    \N{n}(r' > r) \equiv \sum_{k=1}^n \Pins{k}(r' > r).
    \label{eq:nn}
\end{equation}
The expectation values for the powers of the radius $\ev{r^\alpha}_n$, $\alpha > 0$, at the $n$-th injected sphere is obtained by integration over the insertion probability $- \dv*{\Pins{n}(r' > r)}{r}$
\begin{subequations}
    \begin{align}
        \ev{r^\alpha}_n
        & \equiv -\int_0^\infty r^\alpha \dv{r} \Pins{n}(r' > r) \dd{r} \\
        &= \alpha \int_0^\infty r^{\alpha-1} \Pins{n}(r' > r) \dd{r}.
        \label{eq:r_alpha}
    \end{align}
\end{subequations}
The remaining pore space $\pore{n}$ after the $n$-th injection depends, in principle, on the actual realization of this mechanism.
However, in the following, we approximate it by its expectation value
\begin{equation}
    \pore{n} = \vtot - \unit \sum_{k=1}^n \ev{r^d}_n,
\end{equation}
with $\vtot$ the total available volume and $\unit$ the volume of a unit sphere in $d$ dimensions.

\section{Large n limit and fractal dimension}
\label{sec:large-n}

Since the RAP presents a fractal behavior, we \emph{postulate} that the moments $M_\alpha(n)$ describing the radius distribution can be approximated by power-laws at large insertion number $n$
\begin{equation}
    M_\alpha(n) \equiv \sum_{k=1}^n \ev{r^\alpha}_k \underset{n\rightarrow \infty}{\approx} m_\alpha n^{\lambda_\alpha}
\end{equation}
with unknown coefficients $(m_\alpha, \lambda_\alpha)$ and $\alpha$ a real number in $[0, d[$.
Furthermore, we also \emph{postulate} that the pore space converges to $0$ according to a power-law
\begin{equation}
    \pore{n} \underset{n\rightarrow \infty}{\approx} m_d V_d n^{\lambda_d},
\end{equation}
with $\lambda_d < 0$.
We discuss the validity of these two postulates in \cref{sec:numerical}.

With this convention, the expectation value of $r^\alpha$ for $\alpha \in [0,d]$ is
\begin{equation}
    \ev{r^\alpha}_n \underset{n\rightarrow \infty}{\approx} m_\alpha \abs{\lambda_\alpha} n^{\lambda_\alpha -1}.
\end{equation}
In particular, $M_0(n)$ is the number of inserted spheres and $(m_0, \lambda_0) = (1, 1)$.

Given a functional form for $\surface{n}(r)$, we argue that the fractal properties of the RAP are completely determined.
We make different prescriptions for $\surface{n}(r)$ in \cref{sec:ud,sec:it}.
To keep the discussion general, we only assume in this section that $\surface{n}(r)$ is a polynomial
\begin{equation}
    \surface{n}(r) = \sum_{\alpha \in \myset} s_{\alpha}(n) r^{\alpha},
    \label{eq:poly}
\end{equation}
with $\myset$ a discrete set of numbers, not necessarily integers, in the range $[0, d-1]$.

By construction, the insertion probability is normalized
\begin{equation}
    -\int_0^\infty \dv{r} \Pins{n+1}(r' > r) \dd{r} = 1,
\end{equation}
which, upon using \cref{eq:ode,eq:r_alpha,eq:poly}, establishes
\begin{equation}
    \forall n, \quad \pore{n} = \sum_{\alpha \in \myset} \frac{s_{\alpha}(n)}{\alpha + 1} \ev{r^{\alpha + 1}}_{n+1}.
    \label{eq:proba}
\end{equation}

In the large $n$ limit, a change of variable in \cref{eq:r_alpha} absorbs the dependence on $n$
\begin{equation}
    m_\alpha \abs{\lambda_\alpha} = \alpha \int_0^\infty \Pins{n}\qty[r' > n^{(\lambda_\alpha - 1)/ \alpha}x] x^{\alpha - 1} \dd{x}.
    \label{eq:other}
\end{equation}
Since this equation is valid for any $n$ sufficiently large, we \emph{postulate} that the dependence on $n$ vanishes exactly in the argument of $\Pins{n}$.
Therefore
\begin{equation}
    f(x, \alpha) \underset{n\rightarrow \infty}{\equiv} \Pins{n}\qty[r' > n^{(\lambda_\alpha - 1)/ \alpha}x]
    \label{eq:scaling}
\end{equation}
with $f$ function of only one argument.
Similarly, the cumulative radius distribution of \cref{eq:nn} can be expressed in terms of $f$.
\begin{equation}
        \N{n}(r' > r) \underset{n\rightarrow \infty}{\approx} \sum_{k=1}^n f\qty[r k^{(1-\lambda_\alpha)/\alpha}, \alpha].
\end{equation}
The fractal dimension $\fractal$ of the RAP, defined using the slope of the radius cumulative distribution at large $n$, is therefore related to the different $\lambda_\alpha$s according to
\begin{equation}
    \fractal - 1 \equiv \lim_{n\rightarrow \infty} \dv{\ln \N{n}(r' > r)}{\ln r} = \frac{\alpha}{1-\lambda_\alpha}.
    \label{eq:fractal}
\end{equation}
Consequently, the power-law exponents can all be expressed in terms of a single exponent, that we arbitrarily chose to be $\lambda_1$
\begin{equation}
    \lambda_\alpha = \alpha \lambda_1 - (\alpha - 1).
    \label{eq:magic}
\end{equation}
Note that this is consistent with the rather crude approximation that $\ev{r^\alpha}_n \underset{n\rightarrow \infty}{\propto} \ev{r}_n^\alpha$.

\begin{table*}
    \begin{tabular}{c | c c c | c c c | c c c}
        & \multicolumn{3}{c}{Uniform distribution} & \multicolumn{3}{c}{Identical twins} & \multicolumn{3}{c}{Simulations} \\
        \hline
        Exponent & $d=2$ & $d=3$ & $d=4$ & $d=2$ & $d=3$ & $d=4$ & $d=2$ & $d=3$ & $d=4$ \\
        \hline
        \hline
        $\lambda_1$ & $ 0.3789 $ & $ 0.6313 $ & $ 0.7369 $ & $ 0.3614 $ & $ 0.6285 $ & $ 0.7362 $ & $ 0.361 $ & $ 0.634 $ & $ 0.738$ \\
        $\lambda_2$ & $ -0.2421 $ & $ 0.2625 $ & $ 0.4737 $ & $ -0.2771 $ & $ 0.2571 $ & $ 0.4724 $ & $ -0.2778$ & $ 0.265 $ & $ 0.476$ \\
        $\lambda_3$ & -- & $ -0.1062 $ & $ 0.2106 $ & -- & $ -0.1144 $ & $ 0.2086 $ & -- & $ -0.097 $ & $ 0.207$ \\
        $\lambda_4$ & -- & -- & $ -0.0526 $ & -- & -- & $ -0.0552 $ & -- & -- & $ -0.044$ \\
        $\fractal$ & $ 2.6101 $ & $ 3.7119 $ & $ 4.8002 $ & $ 2.5660 $ & $ 3.6921 $ & $ 4.7909 $ &  -- & -- & --
    \end{tabular}
    \caption{On the left columns, the power-law exponents predicted by the mean-field theory. On the right columns, the power-law exponents directly measured over an ensemble average of $256$ simulations of $2 \times 10^7$ spheres each.
    Uncertainties on the determination of the $\lambda_i$s from simulations are in \cref{fig:fit}.
    As explained in \cref{sec:numerical}, we do not provide a direct estimate of the fractal dimension from the simulation.}
    \label{tab:values}
\end{table*}

\section{Surface model I: uniform distribution}
\label{sec:ud}

It should be clear from the previous section that the fractal dimension and the power-law exponents are determined by the yet unspecified function $\surface{n}(r)$.
Our first attempt to model this function is $\surface{n}^{(1)}(r)$, where we make the assumption that all the $n$ first spheres are \emph{uniformly distributed} across the available volume $\vtot$.
In which case, $\surface{n}^{(1)}(r)$ sums the surfaces of spheres with radius $(r + r')$ centered around the $n$ preexisting spheres having radius $r'$ with probability $-\dv*{\N{n}(r' > r)}{r}$
\begin{equation}
    \surface{n}^{(1)}(r) \equiv - d . \unit \int_0^\infty (r+r')^{d-1} \dv{\N{n}}{r'}\dd{r'}.
\end{equation}
In this first model, the set $\myset$ contains all the integers between $0$ and $d-1$, and
\begin{equation}
    s_k(n) = d . \unit \binomial{k}{d-1} M_{d-k-1}(n),
\end{equation}
with $\binomial{k}{n}$ the binomial coefficients.
Given a functional form for $\surface{n}(r)$, \cref{eq:proba,eq:other} form a closed set of equations in the large $n$ limit.
As an example, we explicitly show the computation for $d=2$ in this section, and we direct the interested reader to \cref{app:uniform} for $d=3,4$.

In two dimensions, the ``surface'' $\surface{n}(r)$ is the sum of perimeters at a distance $r$ from the circles of radius $r'$
\begin{equation}
    \surface{n}^{(1)}(r) = - 2 \pi \int_0^\infty (r+r') \dv{\N{n}}{r'}(r') \dd{r'}.
\end{equation}
Therefore $\myset = \{0, 1\}$ and the coefficients $s_\alpha(n)$ in the large $n$ limit are
\begin{subequations}
    \begin{align}
        s_0(n) &= 2 M_1(n) \pi \approx 2 m_1 \pi n^{\lambda_1} \\
        s_1(n) &= 2 M_0(n) \pi \approx 2 \pi n.
    \end{align}
\end{subequations}
\cref{eq:proba}, the asymptotic limit yields
\begin{subequations}
    \begin{align}
        m_2 \pi n^{\lambda_2}
        &= 2 \pi m_1 n^{\lambda_1} \ev{r}_n
        + \pi n \ev{r^2}_n \\
        \implies m_2 n^{\lambda_2} &= 2 m_1^2 \lambda_1 n^{2\lambda_1 - 1}
        - m_2 \lambda_2 n^{\lambda_2}.
    \end{align}
\end{subequations}
As expected, \cref{eq:magic} cancels the dependence on $n$, and we obtain $m_2 = m_1^2$.
Finally, setting $\alpha = 1$ in \cref{eq:other} gives a closed-form equation for $\lambda_1$
\begin{equation}
    \lambda_1 = \int_0^\infty \exp\qty(- 2 x - x^2) \dd{x} = \frac{e \sqrt{\pi}}{2} \erfc(1),
\end{equation}
and an estimate for the fractal dimension
\begin{equation}
    \fractal = 1 + \frac{1}{1-\lambda_1} = 1 + \frac{2}{2-e\sqrt{\pi} \erfc{1}} \approx 2.610.
\end{equation}

Exceptionally, we can find the fractal dimension analytically in two dimensions, but this is not the case in higher dimensions.
As detailed in \cref{app:uniform}, in $d=3$ and $4$, we resort to solving the closed set of equations numerically using a root-finding algorithm.
Predictions for the $\lambda_i$s and the fractal dimension $\fractal$ in $d=2,3$ and $4$ are  collected in \cref{tab:values}.

\section{Surface model II: identical twins}
\label{sec:it}

\begin{figure}
    \includegraphics[width=.4\textwidth]{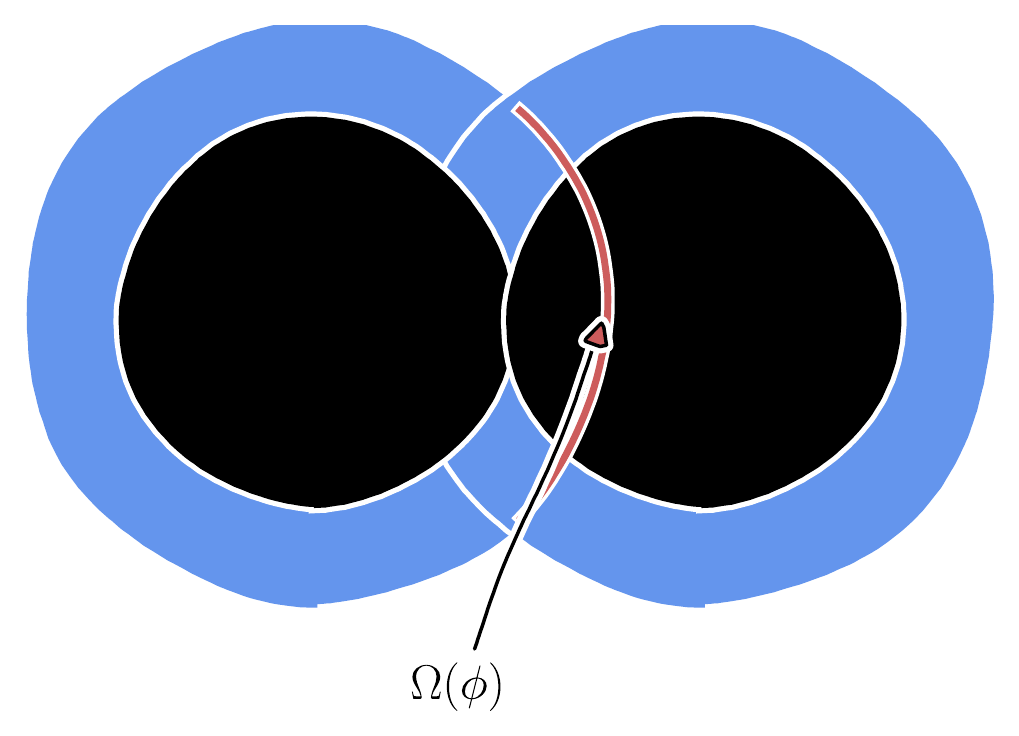}
    \caption{
        We present in \cref{sec:it} a refined model for the ``surface'' function $\surface{n}^{(2)}(r)$.
        Each sphere having necessarily at least one neighbor, spheres cannot nucleate in a fraction of the surface $\surface{n}(r)$.
        We estimate this fraction by the hyperspherical cap $\Omega(\phi)$ truncated when two spheres of equal radius collide.
    }
    \label{fig:picture_2}
\end{figure}

In a refined attempt to model the ``surface'' function $\surface{n}^{(2)}(r)$, we approximate the \emph{a priori} complex network of sphere collisions by the first order correction due to having one collision with a sphere of identical radius.
In other words, we consider isolated pairs of ``identical twins'' to account for the close-range effect of having at least one neighbor for each sphere.
We illustrate it in \cref{fig:picture_2}.

Let us have two spheres of radius $r'$ in $d$ dimensions tangent to one another, the ``surface'' located at a distance $r$ from the two spheres is truncated by $2 \Omega(\phi)$, the area of the unit hyperspherical cap of half-angle at the summit $\phi = \arccos[r'/(r+r')]$
\begin{equation}
    \Omega(\phi) = \frac{d \unit}{2} I_{\sin^2\phi}\qty(\frac{d-1}{2},\frac{1}{2}).
\end{equation}
$I$ is the regularized incomplete beta function.
The presence of this ``identical twin'' for small $r$ modifies the shape of the surface function $\surface{n}(r)$
\begin{align}
    \surface{n}^{(2)}(r) &\equiv \surface{n}^{(1)} - \frac{(2 \pi r)^{\frac{d-1}{2}}}{\Gamma \left(\frac{d+1}{2}\right)} M_{\frac{d-1}{2}}(n).
    \label{eq:twins}
\end{align}
For even dimensions, the surface function now includes half-integer powers of $r$.

As in the previous section, we compute the power-law exponents in two dimensions and give the higher dimensional derivation in \cref{app:twins}.
The values predicted for the power-law exponents are collected in \cref{tab:values}.
For $d=2$, the surface function is a sum of three terms with $\myset = \{0, 1/2, 1\}$
\begin{subequations}
    \begin{align}
        s_0(n) &= 2 M_1(n) \pi = 2 m_1 \pi n^{\lambda_1}, \\
        s_{1/2}(n) &= - 2 \sqrt{2} M_{1/2}(n) = - 2 \sqrt{2} m_{1/2} n^{\lambda_{1/2}}, \\
        s_1(n) &= 2 M_0(n) \pi = 2 \pi n.
    \end{align}
\end{subequations}
\cref{eq:other} for $\alpha \in \{1/2, 1, 2\}$ forms a closed-system of three equations
\begin{widetext}
    \begin{subequations}
        \begin{align}
            \frac{m_{1 / 2}}{m_1^{1/2}} \lambda_{1 / 2} &= \frac{1}{2} \int_0^\infty \frac{1}{\sqrt{x}} \exp[- \frac{m_1^2}{m_2} \qty(x^2 + 2 x - \frac{4\sqrt{2}}{3\pi} \frac{m_{1/2}}{m_1^{1/2}} x^{3/2})] \dd{x}, \\
            \lambda_1 &= \int_0^\infty \exp[- \frac{m_1^2}{m_2} \qty(x^2 + 2 x - \frac{4\sqrt{2}}{3\pi} \frac{m_{1/2}}{m_1^{1/2}} x^{3/2})] \dd{x}, \\
            - \frac{m_2}{m_1^2} \lambda_2 &= 2 \int_0^\infty x \exp[- \frac{m_1^2}{m_2} \qty(x^2 + 2 x - \frac{4\sqrt{2}}{3\pi} \frac{m_{1/2}}{m_1^{1/2}} x^{3/2})] \dd{x}.
        \end{align}
    \end{subequations}
\end{widetext}
with three unknowns $\lambda_1, m_{1/2}/m_1^{1/2}, m_2/m_1^2$.
This system can be solved with a root finding algorithm, and we find
\begin{equation*}
    \lambda_1 \approx 0.3614, \quad
    \frac{m_{1/2}}{m_1^{1/2}} \approx 0.7981, \quad
    \frac{m_2}{m_1^2} \approx 0.8188.
\end{equation*}

\begin{figure}
    \includegraphics[width=.4\textwidth]{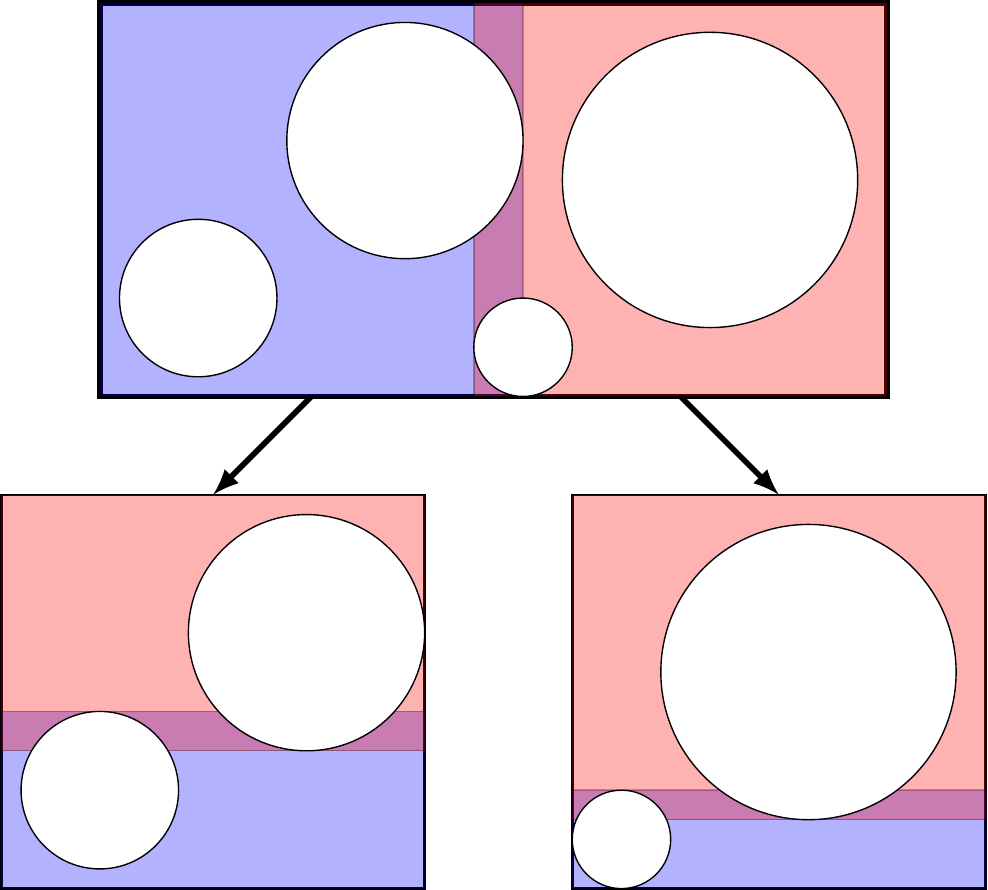}
    \caption{
        Illustration of the space partitioning method used in the numerical simulation.
        Each node of the tree consists of a set of spheres and a bounding box.
        After the insertion of $\order{100}$ spheres, the node is divided into two children nodes and the set of spheres is split into the two children nodes.
        Each of the children's bounding boxes (in red and blue) is designed to enclose at least half of the spheres of the parent node.
        Overlap usually occurs between the bounding boxes of the two children.
        A few spheres may not fit in either of the children nodes, in which case they remain associated with the parent node.
    }
    \label{fig:algo}
\end{figure}

\section{Numerical simulations}
\label{sec:numerical}

\begin{figure}
    \includegraphics[width=.4\textwidth]{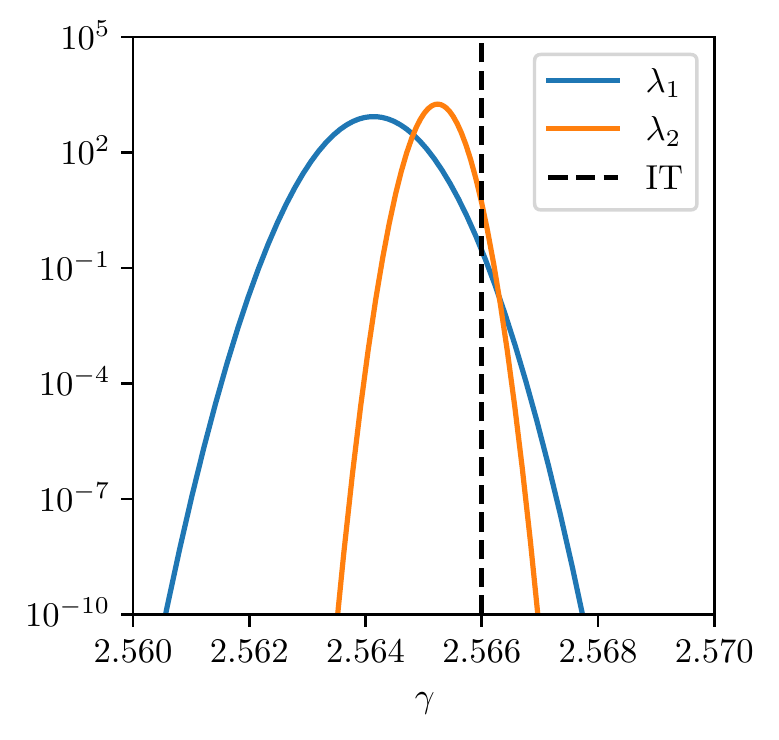}
    \caption{Likelihood for the fractal dimension $\fractal$ estimated from the power-law exponents $\lambda_i$s derived in the simulation using \cref{eq:fractal,eq:magic}. The dashed black line is the prediction from the Identical Twins model.}
    \label{fig:fractal}
\end{figure}

\begin{figure*}
    \subfloat[]{\includegraphics[width=.32\textwidth]{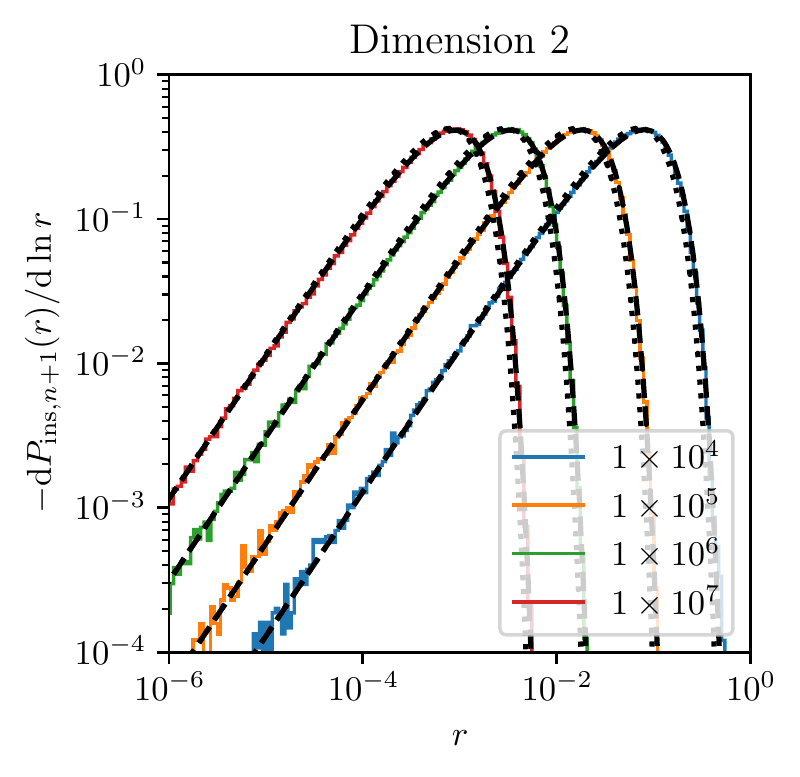}}
    \subfloat[]{\includegraphics[width=.32\textwidth]{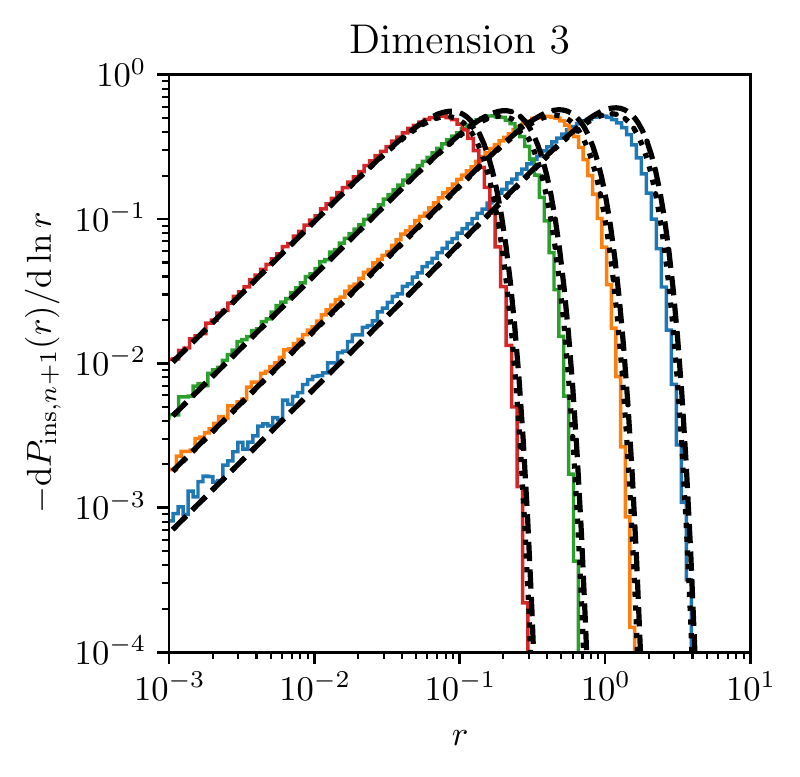}}
    \subfloat[]{\includegraphics[width=.32\textwidth]{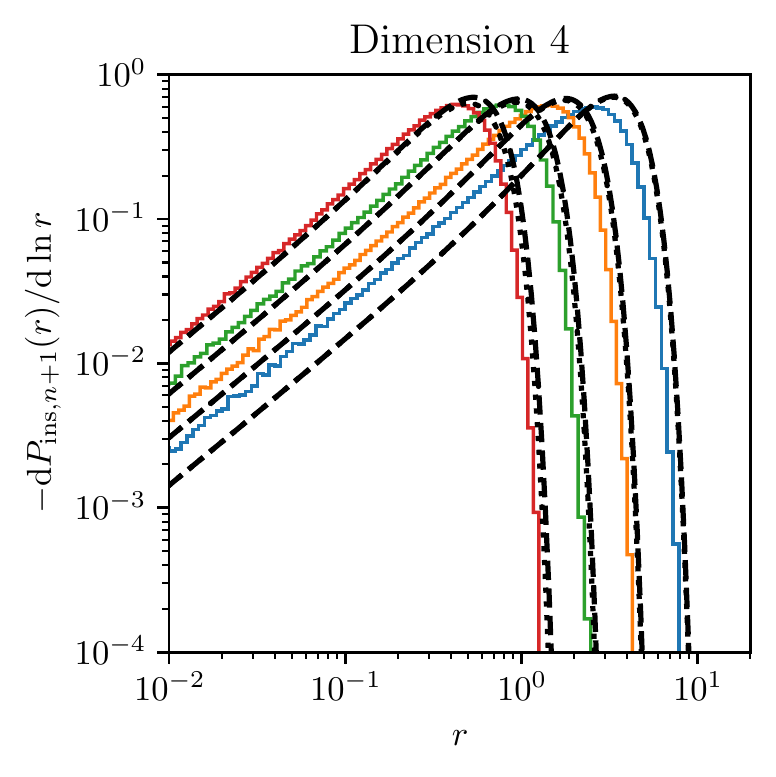}}
    \caption{Insertion probability $-\dv*{\Pins{n+1}}{\ln r}$. The solid colored lines are estimated by attempting $10^6$ nucleations in a preexisting packing of $10^4, 10^5, 10^6$ and $10^7$ spheres.
    The dotted black lines show the prediction from the uniform distribution model of \cref{sec:ud} and the dashed black lines from the identical twins model of \cref{sec:it}.}
    \label{fig:insert}
\end{figure*}

A step in the simulation starts by selecting a random nucleation site inside a $d$-dimensional square box.
The sphere takes the largest radius possible to avoid overlap with other spheres or with the boundaries of the box.
The closest sphere is determined using a space-partitioning method, illustrated in \cref{fig:algo}, which reduces the complexity from $\order{n}$ to $\order{\ln n}$.
If the nucleation site lies inside a preexisting sphere, it is discarded and a new nucleation site is drawn.
As the number of spheres increases, the rejection rate increases thus making the insertion of spheres increasingly challenging, especially in low dimensions.

First, we perform a series of numerical simulation to validate our model for the insertion probability of \cref{eq:master}.
To this end, we attempt $10^6$ test insertions in fixed RAPs of $10^4, 10^5, 10^6$ and $10^7$ spheres.
We compare in \cref{fig:insert} this numerical result with our two models: the Uniform Distribution (UD) and the Identical Twins (IT) models.
Since the realization of the RAP is known and fixed, we use for the $M_\alpha(n)$s the actual radii in the simulation instead of their expectation values.
Overall, we find a good agreement in $d=2$ and $3$ and observe a small discrepancy for $d=4$ that vanishes as $n$ gets larger.
We interpret this deviation at $d=4$ as a boundary effect that we do not account for in the model.
Even after $10$ million insertions, more than half of the simulation box is still empty and the surface of the bounding box is equivalent to the surface of the packed spheres.

For comparison purposes, we review Ref~\cite{PhysRevE.65.056108} where a similar approach was originally proposed.
In this article, the authors did not account for the fact that a fraction of these layers $\surface{n}(r)\delta r$ should be discarded, nor did they account for the growing surface of these layers.
Using our formalism, they assume
\begin{equation*}
    \volume{n}(r + \delta r) = \volume{n}(r) + \surface{n}(0) \delta r.
\end{equation*}
From this equation, they infer that the cumulative insertion probability is affine and impose an artificial cutoff to keep it positive
\begin{equation*}
    \Pins{n+1}(r' > r) = \qty[1 - \frac{\surface{n}(0)}{\pore{n}} r] \Theta\qty[\frac{\pore{n}}{\surface{n}(0)} - r].
\end{equation*}
This insertion probability of Ref~\cite{PhysRevE.65.056108} matches ours and the numerical simulations on small scales but deviates significantly when the radius increases.
In the end, the authors of Ref~\cite{PhysRevE.65.056108} give, as an estimate of the fractal dimension
\begin{equation}
    \fractal \approx d + \frac{d + 1}{d + 2} =
    \begin{cases}
        2.75 & d=2 \\
        3.8 & d=3 \\
        4.83 & d=4
    \end{cases}.
\end{equation}
They find that this formula overestimates the fractal dimension for $d=2$ and $3$ but reproduces well their findings in $d=4$ with a single simulation of $5$ million spheres.

\begin{figure*}
    \begin{tabular}{c c c}
        \subfloat[]{\includegraphics[width=.28\textwidth]{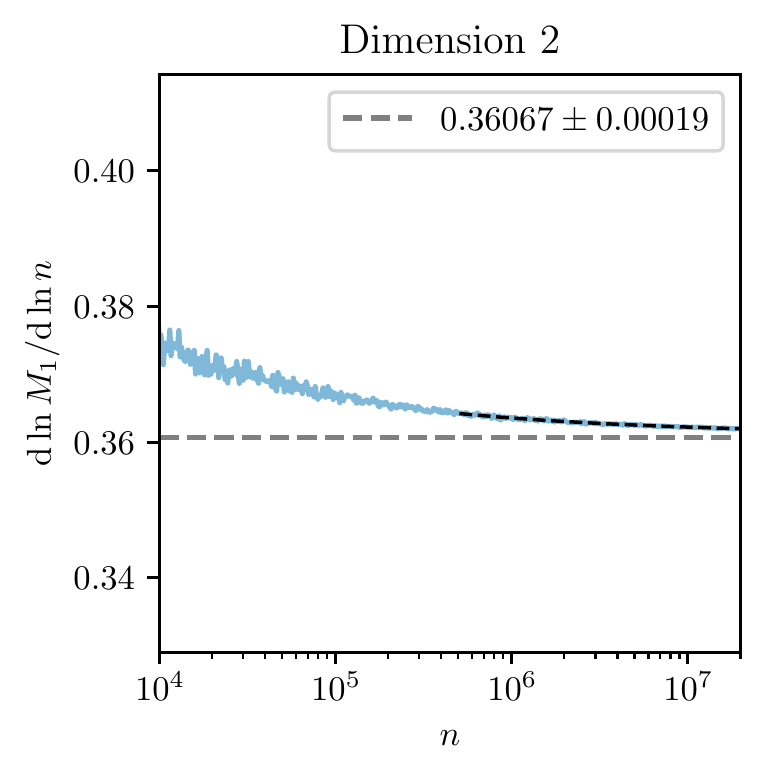}} &
        \subfloat[]{\includegraphics[width=.28\textwidth]{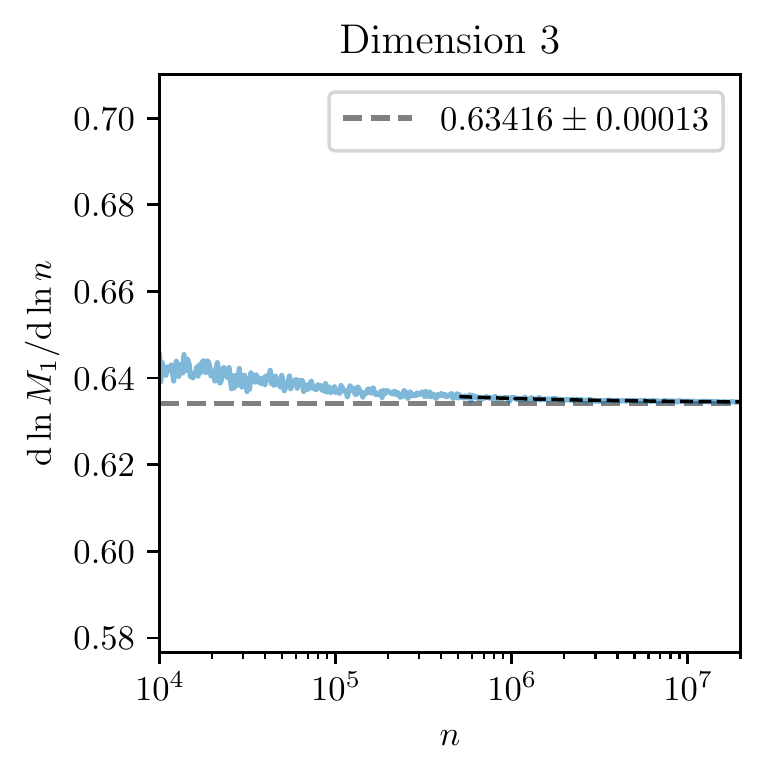}} &
        \subfloat[]{\includegraphics[width=.28\textwidth]{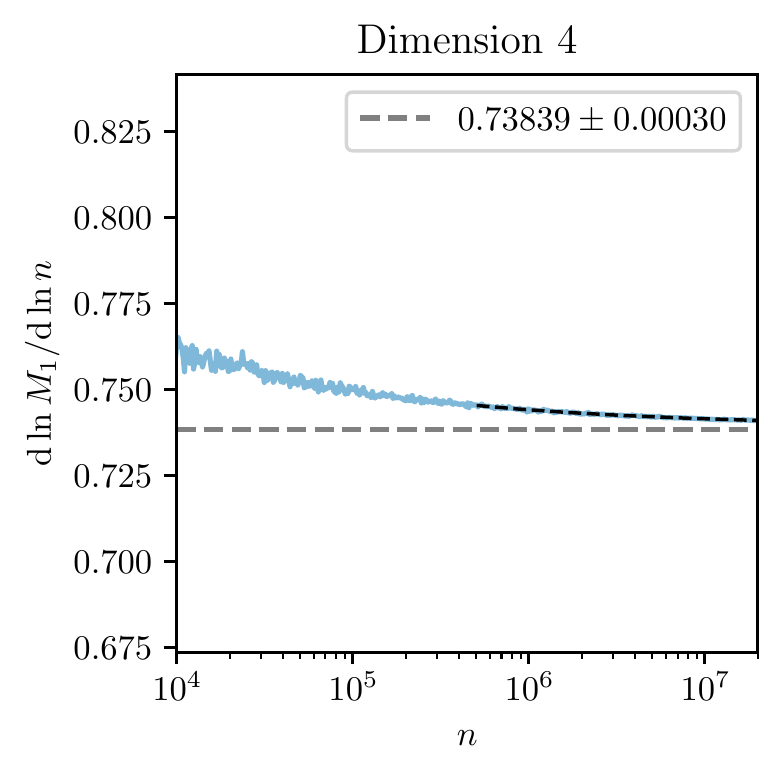}}\\

        &
        \subfloat[]{\includegraphics[width=.28\textwidth]{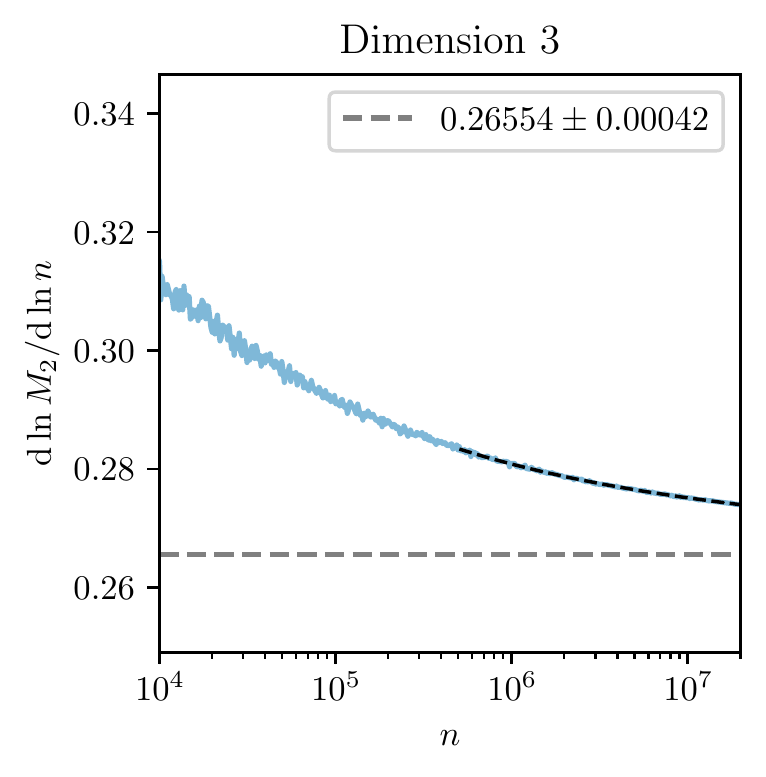}} &
        \subfloat[]{\includegraphics[width=.28\textwidth]{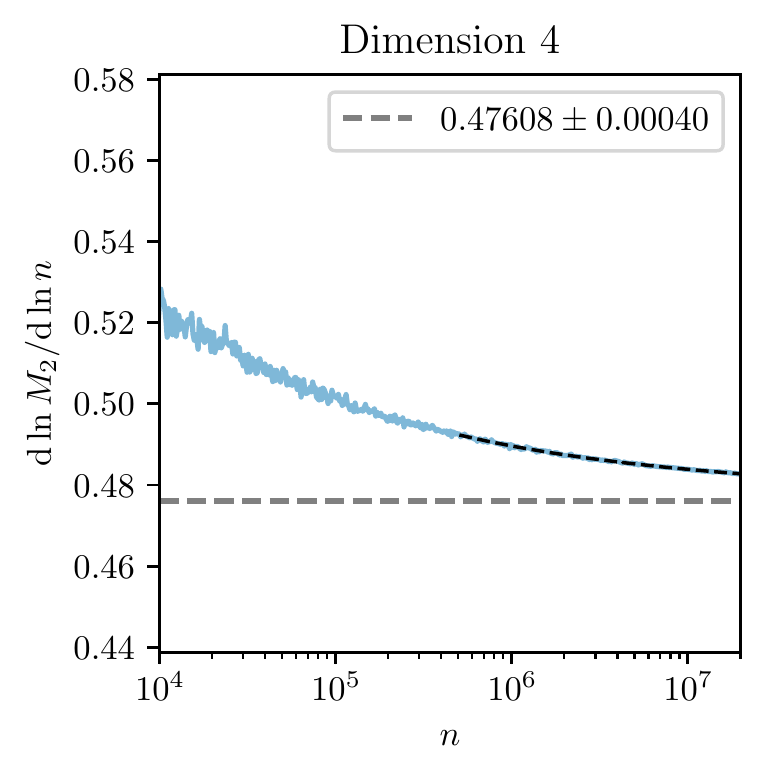}} \\

        &
        &
        \subfloat[]{\includegraphics[width=.28\textwidth]{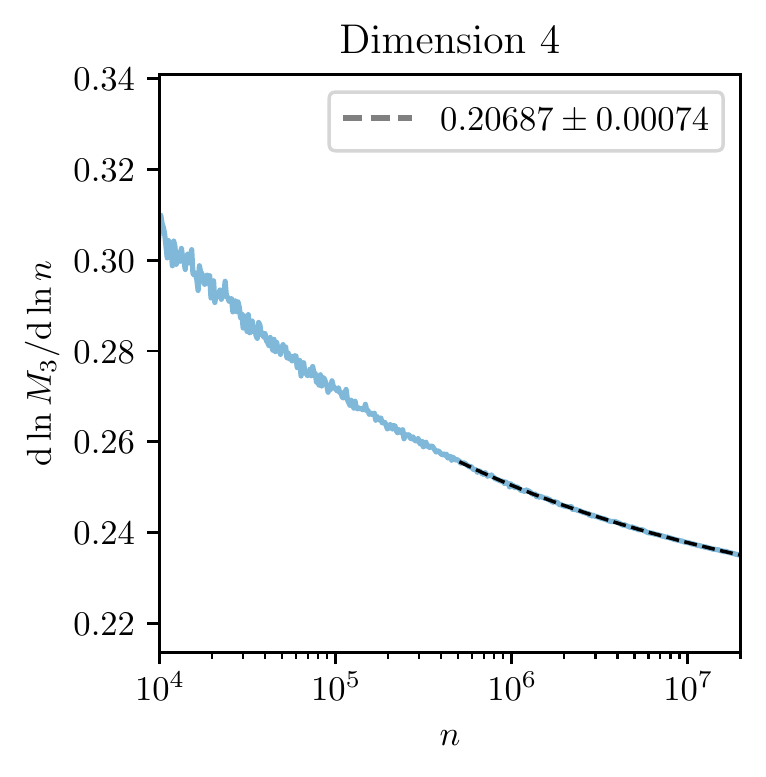}}\\

        \subfloat[]{\includegraphics[width=.28\textwidth]{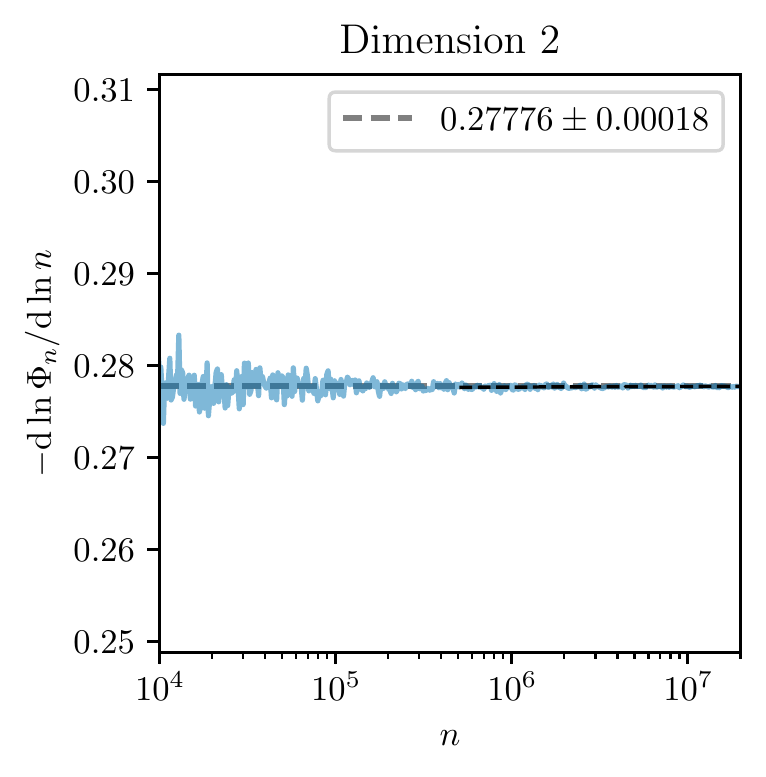}} &
        \subfloat[]{\includegraphics[width=.28\textwidth]{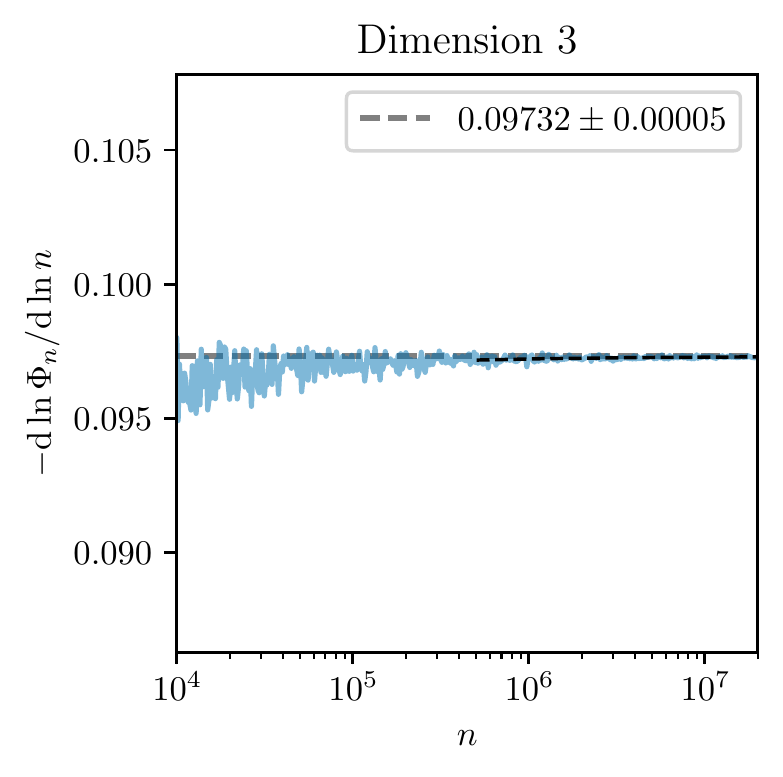}} &
        \subfloat[]{\includegraphics[width=.28\textwidth]{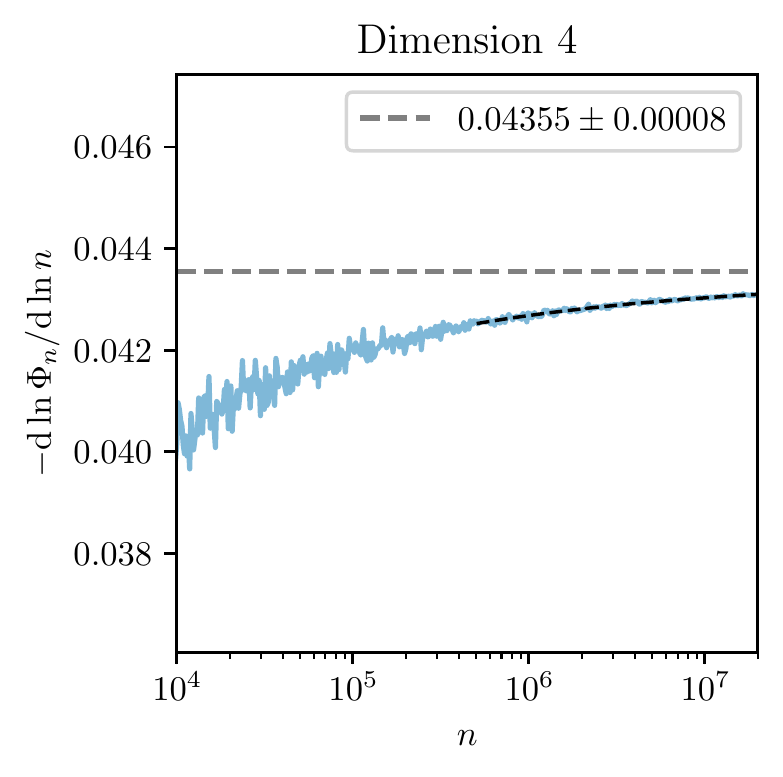}}

    \end{tabular}

    \caption{Estimation of the exponents $\abs{\lambda_i}$ from an ensemble average of $256$ simulations of $2\times 10^7$ spheres. The dashed black curves are the fit to the data accounting for its approach to its asymptotic value as in \cref{eq:asympt}. The dashed gray curves are the asymptotic values found.}
    \label{fig:fit}
\end{figure*}

Second, we make ensemble averages of sets of $256$ numerical simulations, each containing $2\times 10^7$ spheres.
We use this ensemble average to estimate the $\lambda_i$s and validate our postulates that the $M_\alpha(n)$ approach a power-law behavior.
As can be seen in \cref{fig:fit}, the estimators of the power-law exponents have reached their asymptotic value in two dimensions, but are still approaching their asymptote in higher dimensions.
To account for the late convergence of our estimators, the $\lambda_i$s are fitted together with their approach to the asymptotic value
\begin{equation}
    \dv{\ln M_i}{\ln n} \approx \lambda_i + b (\ln n)^c.
    \label{eq:asympt}
\end{equation}
The $\lambda_i$s we infer from the numerical simulations are broadly consistent with Ref~\cite{PhysRevE.65.056108}.
However, we ask the reader to take these fitting values with care for $d=3$ and particularly for $d=4$ when comparing with our model.
Indeed, it is clear that these simulations have not yet reached the fractal regime and are subject to boundary effects.
The authors of Ref~\cite{PhysRevE.65.056108} already encountered the same issue when giving estimates for the fractal dimension $\gamma$.

Finally, we give in \cref{fig:fractal-sim} a direct estimation of the fractal dimension from the cumulative radius distribution.
As already mentioned, the cumulative radius distribution does not yet present a power-law behavior for $d=3$ and $4$, thus making hazardous a direct determination of the fractal dimension.
Nonetheless, we test our model in two dimensions by performing an indirect estimation of the fractal dimension $\fractal$ using each of the best fit for $\lambda_1$ and $\lambda_2$.
To do so, we use \cref{eq:fractal,eq:magic} and assume that the error on the measured $\lambda_i$s is gaussian
The resulting likelihood distributions for the fractal dimension $\fractal$ is shown in \cref{fig:fractal}, together with the prediction from the identical twins model.
The predicted value for $\gamma$ is also given as a horizontal line in \cref{fig:fractal-sim} and show perfect agreement with the direct estimation of $\gamma$.

\section{Discussion}

In this article, we presented a ``mean-field inspired'' model to understand the properties of Random Apollonian Packings, a prototypal example of sphere packing.
This model gives a prediction for the insertion probability with unprecedented support by simulations in $d=2, 3$ and, to a smaller extent, in $d=4$.
The agreement between the model and the simulation weakens the higher the dimension and the lower the number of inserted spheres.
This is to be expected since, in this regime the available space $\vtot$ is rather empty and boundary effects remain important, hence our mean-field approximation breaks down.

Then, we presented a systematic method to determine the asymptotic behavior of the moments $M_\alpha(n)$ directly from the insertion probability without assuming \emph{a priori} a functional form for the radius distribution $\N{n}(r)$.
We find very good agreement in two dimensions, however for $d=3$ and $4$, our simulations have clearly not yet reach the fractal regime to make an accurate comparison.
Nonetheless, our estimations for the power-law exponents $\lambda_i$s are broadly consistent with values inferred from the simulations.

\begin{figure*}
    \subfloat[]{\includegraphics[width=.28\textwidth]{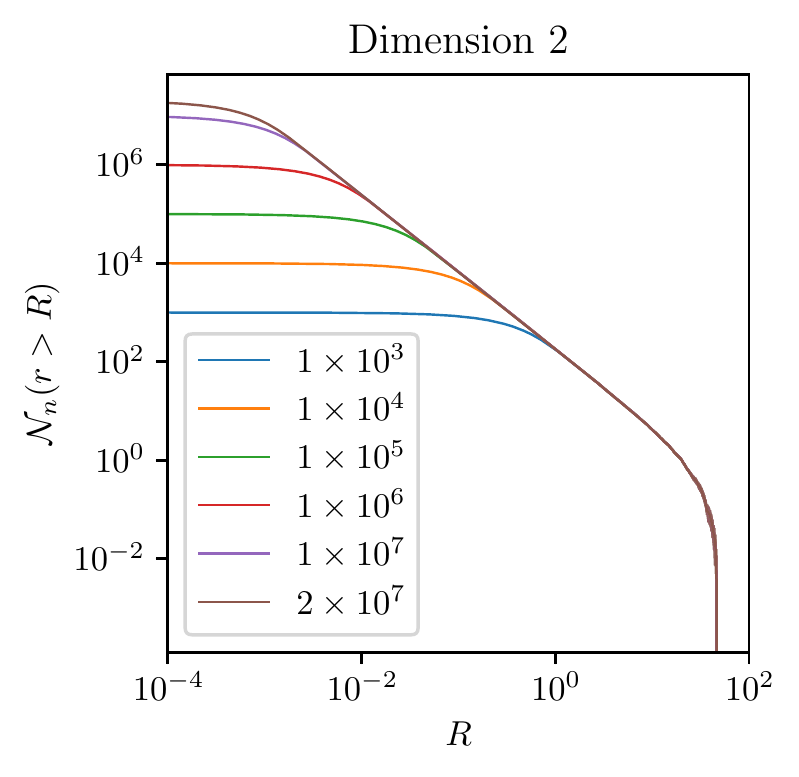}}
    \subfloat[]{\includegraphics[width=.28\textwidth]{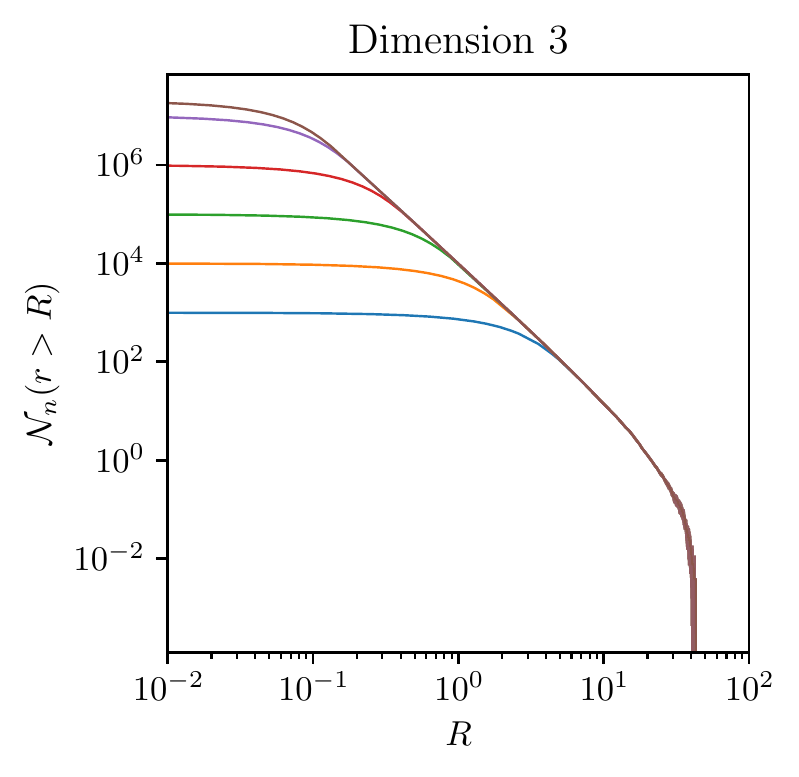}}
    \subfloat[]{\includegraphics[width=.28\textwidth]{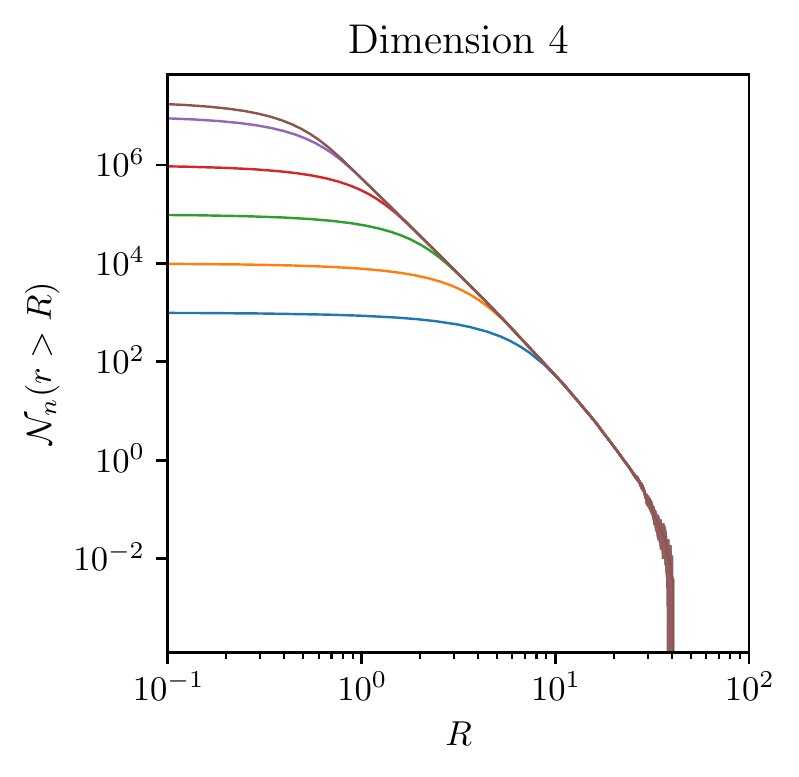}}

    \subfloat[]{\includegraphics[width=.28\textwidth]{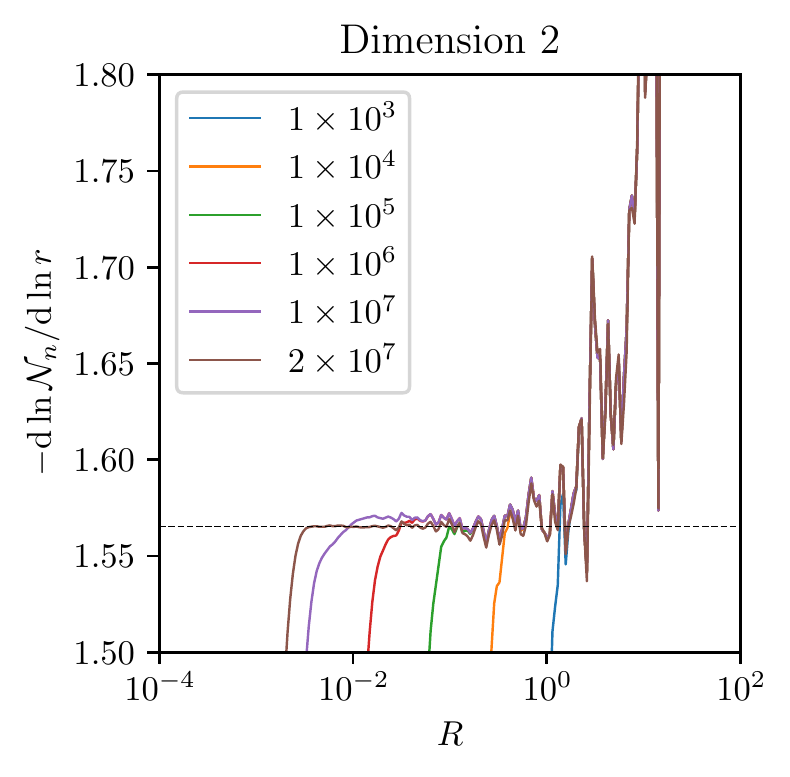}}
    \subfloat[]{\includegraphics[width=.28\textwidth]{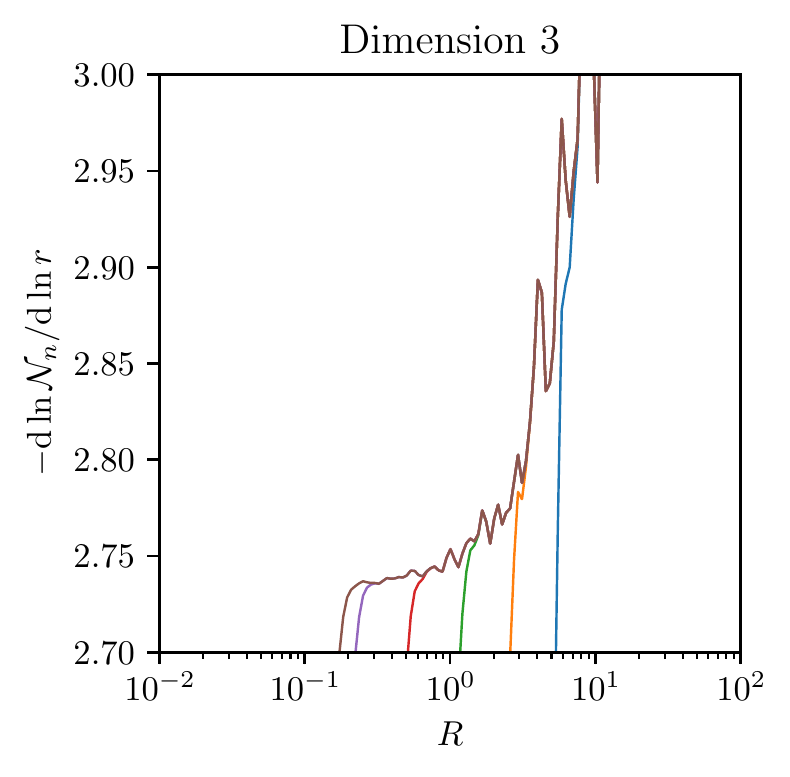}}
    \subfloat[]{\includegraphics[width=.28\textwidth]{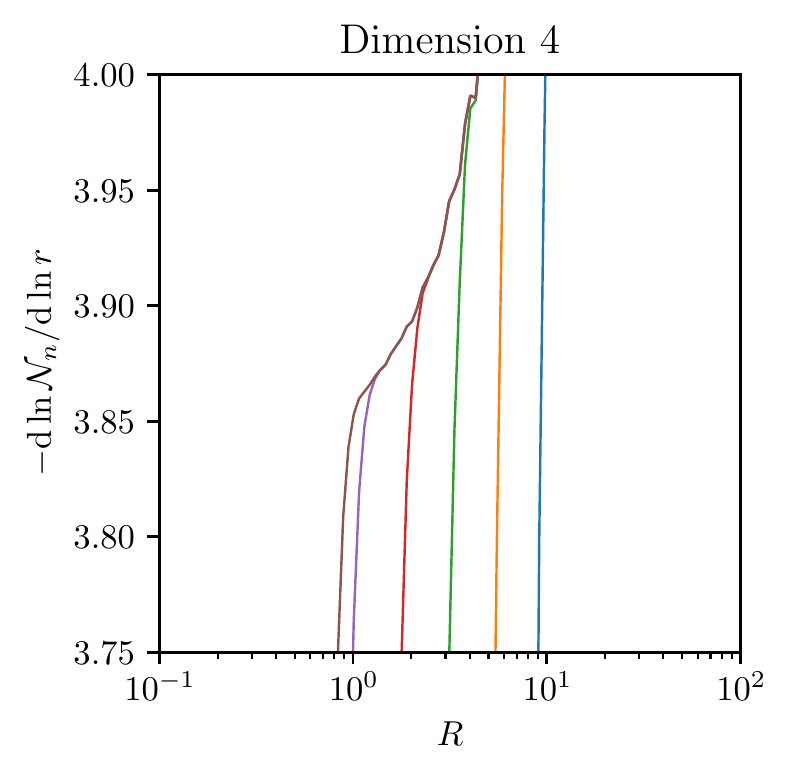}}
    \caption{
    Upper panels (a)-(c): Ensemble average of the cumulative distribution function $\N{n}(r>R)$ for a RAP into a $d$-dimensional square box of size $100$ in arbitrary units, obtained by averaging over $256$ independent realizations.
    Lower panels (d)-(f): Estimation of the fractal dimension $\fractal$ from an ensemble average of $256$ simulations of $2\times 10^7$ spheres. Differentation is done using a central difference scheme.}
    \label{fig:fractal-sim}
\end{figure*}

We stress that the simulations in $4$ dimensions are still far from the expected power-law behavior, as can be assessed in \cref{fig:fit,fig:fractal-sim}.
Indeed, we observe that less than half the available volume is occupied by spheres and that the surface of the bounding box is of the same order as the surface of the spheres even after $20$ million insertions.
It is unclear whether our best fits for the power-law exponents reflect the fractal regime or a transient regime suffering from a boundary effect.

In summary, the method presented in this article gives us an analytical handle on both the \emph{global} scaling properties of this "Packing-Limited Growth" problem and the \emph{step-by-step} properties in the form of the insertion probability.

In the future, we plan to apply this mean-field approach to broader classes of "Packing-Limited Growth" problems, such as extensions with nonspherical particles. We also plan to investigate boundary effects in higher dimensions, by allowing different boundary topologies and by running simulations with an even larger number of spheres.

\begin{acknowledgments}
It is a pleasure to thank Christophe Ringeval and Dani\`ele Steer for their support and encouragements.
This work is partially supported by the Wallonia-Brussels Federation Grant ARC \textnumero~19/24-103.

\end{acknowledgments}

\bibliography{biblio}

\begin{thebibliography}{17}
\expandafter\ifx\csname natexlab\endcsname\relax\def\natexlab#1{#1}\fi
\expandafter\ifx\csname bibnamefont\endcsname\relax
  \def\bibnamefont#1{#1}\fi
\expandafter\ifx\csname bibfnamefont\endcsname\relax
  \def\bibfnamefont#1{#1}\fi
\expandafter\ifx\csname citenamefont\endcsname\relax
  \def\citenamefont#1{#1}\fi
\expandafter\ifx\csname url\endcsname\relax
  \def\url#1{\texttt{#1}}\fi
\expandafter\ifx\csname urlprefix\endcsname\relax\def\urlprefix{URL }\fi
\providecommand{\bibinfo}[2]{#2}
\providecommand{\eprint}[2][]{\url{#2}}

\bibitem[{\citenamefont{Horn}(1971)}]{horn1971adaptive}
\bibinfo{author}{\bibfnamefont{H.~S.} \bibnamefont{Horn}},
  \emph{\bibinfo{title}{The adaptive geometry of trees}}
  (\bibinfo{publisher}{Princeton University Press}, \bibinfo{year}{1971}).

\bibitem[{\citenamefont{Van~der Marck}(1996)}]{van1996network}
\bibinfo{author}{\bibfnamefont{S.}~\bibnamefont{Van~der Marck}},
  \bibinfo{journal}{Physical review letters} \textbf{\bibinfo{volume}{77}},
  \bibinfo{pages}{1785} (\bibinfo{year}{1996}).

\bibitem[{\citenamefont{Conway and Sloane}(2013)}]{conway2013sphere}
\bibinfo{author}{\bibfnamefont{J.~H.} \bibnamefont{Conway}} \bibnamefont{and}
  \bibinfo{author}{\bibfnamefont{N.~J.~A.} \bibnamefont{Sloane}},
  \emph{\bibinfo{title}{Sphere packings, lattices and groups}}, vol.
  \bibinfo{volume}{290} (\bibinfo{publisher}{Springer Science \& Business
  Media}, \bibinfo{year}{2013}).

\bibitem[{\citenamefont{Gaite and Manrubia}(2002)}]{Gaite:2002mq}
\bibinfo{author}{\bibfnamefont{J.}~\bibnamefont{Gaite}} \bibnamefont{and}
  \bibinfo{author}{\bibfnamefont{S.~C.} \bibnamefont{Manrubia}},
  \bibinfo{journal}{Monthly Notices of the Royal Astronomical Society}
  \textbf{\bibinfo{volume}{335}}, \bibinfo{pages}{977} (\bibinfo{year}{2002}),
  ISSN \bibinfo{issn}{0035-8711, 1365-2966}, \eprint{astro-ph/0205188}.

\bibitem[{\citenamefont{Gaite}(2005)}]{Gaite:2005di}
\bibinfo{author}{\bibfnamefont{J.}~\bibnamefont{Gaite}}, \bibinfo{journal}{The
  European Physical Journal B} \textbf{\bibinfo{volume}{47}},
  \bibinfo{pages}{93} (\bibinfo{year}{2005}), ISSN \bibinfo{issn}{1434-6028,
  1434-6036}, \eprint{astro-ph/0506543}.

\bibitem[{\citenamefont{Gaite}(2006)}]{Gaite:2006tr}
\bibinfo{author}{\bibfnamefont{J.}~\bibnamefont{Gaite}},
  \bibinfo{journal}{Physica D: Nonlinear Phenomena}
  \textbf{\bibinfo{volume}{223}}, \bibinfo{pages}{248} (\bibinfo{year}{2006}),
  ISSN \bibinfo{issn}{01672789}, \eprint{astro-ph/0603572}.

\bibitem[{\citenamefont{Borrill et~al.}(1995)\citenamefont{Borrill, Kibble,
  Vachaspati, and Vilenkin}}]{Borrill:1995gu}
\bibinfo{author}{\bibfnamefont{J.}~\bibnamefont{Borrill}},
  \bibinfo{author}{\bibfnamefont{T.~W.~B.} \bibnamefont{Kibble}},
  \bibinfo{author}{\bibfnamefont{T.}~\bibnamefont{Vachaspati}},
  \bibnamefont{and} \bibinfo{author}{\bibfnamefont{A.}~\bibnamefont{Vilenkin}},
  \bibinfo{journal}{Phys. Rev. D} \textbf{\bibinfo{volume}{52}},
  \bibinfo{pages}{1934} (\bibinfo{year}{1995}), \eprint{hep-ph/9503223}.

\bibitem[{\citenamefont{Kosowsky et~al.}(1992)\citenamefont{Kosowsky, Turner,
  and Watkins}}]{Kosowsky:1991ua}
\bibinfo{author}{\bibfnamefont{A.}~\bibnamefont{Kosowsky}},
  \bibinfo{author}{\bibfnamefont{M.~S.} \bibnamefont{Turner}},
  \bibnamefont{and} \bibinfo{author}{\bibfnamefont{R.}~\bibnamefont{Watkins}},
  \bibinfo{journal}{Phys. Rev. D} \textbf{\bibinfo{volume}{45}},
  \bibinfo{pages}{4514} (\bibinfo{year}{1992}).

\bibitem[{\citenamefont{Auclair et~al.}(2022)\citenamefont{Auclair, Caprini,
  Cutting, Hindmarsh, Rummukainen, Steer, and Weir}}]{Auclair:2022jod}
\bibinfo{author}{\bibfnamefont{P.}~\bibnamefont{Auclair}},
  \bibinfo{author}{\bibfnamefont{C.}~\bibnamefont{Caprini}},
  \bibinfo{author}{\bibfnamefont{D.}~\bibnamefont{Cutting}},
  \bibinfo{author}{\bibfnamefont{M.}~\bibnamefont{Hindmarsh}},
  \bibinfo{author}{\bibfnamefont{K.}~\bibnamefont{Rummukainen}},
  \bibinfo{author}{\bibfnamefont{D.~A.} \bibnamefont{Steer}}, \bibnamefont{and}
  \bibinfo{author}{\bibfnamefont{D.~J.} \bibnamefont{Weir}},
  \bibinfo{journal}{JCAP} \textbf{\bibinfo{volume}{09}}, \bibinfo{pages}{029}
  (\bibinfo{year}{2022}), \eprint{2205.02588}.

\bibitem[{\citenamefont{Andrienko et~al.}(1994)\citenamefont{Andrienko,
  Brilliantov, and Krapivsky}}]{andrienko1994pattern}
\bibinfo{author}{\bibfnamefont{Y.~A.} \bibnamefont{Andrienko}},
  \bibinfo{author}{\bibfnamefont{N.}~\bibnamefont{Brilliantov}},
  \bibnamefont{and}
  \bibinfo{author}{\bibfnamefont{P.}~\bibnamefont{Krapivsky}},
  \bibinfo{journal}{Journal of statistical physics}
  \textbf{\bibinfo{volume}{75}}, \bibinfo{pages}{507} (\bibinfo{year}{1994}).

\bibitem[{\citenamefont{Brilliantov et~al.}(1994)\citenamefont{Brilliantov,
  Krapivsky, and Andrienko}}]{brilliantov1994random}
\bibinfo{author}{\bibfnamefont{N.}~\bibnamefont{Brilliantov}},
  \bibinfo{author}{\bibfnamefont{P.}~\bibnamefont{Krapivsky}},
  \bibnamefont{and} \bibinfo{author}{\bibfnamefont{Y.~A.}
  \bibnamefont{Andrienko}}, \bibinfo{journal}{Journal of Physics A:
  Mathematical and General} \textbf{\bibinfo{volume}{27}},
  \bibinfo{pages}{L381} (\bibinfo{year}{1994}).

\bibitem[{\citenamefont{Dodds and Weitz}(2002)}]{PhysRevE.65.056108}
\bibinfo{author}{\bibfnamefont{P.~S.} \bibnamefont{Dodds}} \bibnamefont{and}
  \bibinfo{author}{\bibfnamefont{J.~S.} \bibnamefont{Weitz}},
  \bibinfo{journal}{Phys. Rev. E} \textbf{\bibinfo{volume}{65}},
  \bibinfo{pages}{056108} (\bibinfo{year}{2002}),
  \urlprefix\url{https://link.aps.org/doi/10.1103/PhysRevE.65.056108}.

\bibitem[{\citenamefont{Delaney et~al.}(2008)\citenamefont{Delaney, Hutzler,
  and Aste}}]{delaney2008relation}
\bibinfo{author}{\bibfnamefont{G.~W.} \bibnamefont{Delaney}},
  \bibinfo{author}{\bibfnamefont{S.}~\bibnamefont{Hutzler}}, \bibnamefont{and}
  \bibinfo{author}{\bibfnamefont{T.}~\bibnamefont{Aste}},
  \bibinfo{journal}{Physical review letters} \textbf{\bibinfo{volume}{101}},
  \bibinfo{pages}{120602} (\bibinfo{year}{2008}).

\bibitem[{\citenamefont{Manna}(1992)}]{manna1992space}
\bibinfo{author}{\bibfnamefont{S.}~\bibnamefont{Manna}},
  \bibinfo{journal}{Physica A: Statistical Mechanics and its Applications}
  \textbf{\bibinfo{volume}{187}}, \bibinfo{pages}{373} (\bibinfo{year}{1992}).

\bibitem[{\citenamefont{Manna and Herrmann}(1991)}]{manna1991precise}
\bibinfo{author}{\bibfnamefont{S.}~\bibnamefont{Manna}} \bibnamefont{and}
  \bibinfo{author}{\bibfnamefont{H.}~\bibnamefont{Herrmann}},
  \bibinfo{journal}{Journal of Physics A: Mathematical and General}
  \textbf{\bibinfo{volume}{24}}, \bibinfo{pages}{L481} (\bibinfo{year}{1991}).

\bibitem[{\citenamefont{Mandelbrot and
  Mandelbrot}(1982)}]{mandelbrot1982fractal}
\bibinfo{author}{\bibfnamefont{B.~B.} \bibnamefont{Mandelbrot}}
  \bibnamefont{and} \bibinfo{author}{\bibfnamefont{B.~B.}
  \bibnamefont{Mandelbrot}}, \emph{\bibinfo{title}{The fractal geometry of
  nature}}, vol.~\bibinfo{volume}{1} (\bibinfo{publisher}{WH freeman New York},
  \bibinfo{year}{1982}).

\bibitem[{\citenamefont{Amirjanov and Sobolev}(2012)}]{amirjanov2012fractal}
\bibinfo{author}{\bibfnamefont{A.}~\bibnamefont{Amirjanov}} \bibnamefont{and}
  \bibinfo{author}{\bibfnamefont{K.}~\bibnamefont{Sobolev}},
  \bibinfo{journal}{Advanced Powder Technology} \textbf{\bibinfo{volume}{23}},
  \bibinfo{pages}{591} (\bibinfo{year}{2012}).

\end{thebibliography}

\onecolumngrid

\appendix

\section{Uniform distribution models in higher dimensions}
\label{app:uniform}

\subsection{Dimension $d=3$}
\label{app:d3}

In three dimensions, the surface function of the uniform density model is given by
\begin{equation}
    \surface{n}^{(1)}(r) = -4 \pi \int_0^\infty (r+r')^2 \dv{\N{n}}{r'} \dd{r'}
\end{equation}
Therefore $\myset = \{0, 1, 2\}$ and
\begin{subequations}
    \begin{align}
        s_0(n) &= 4 \pi M_2(n) = 4 m_2 \pi n^{\lambda_2}\\
        s_1(n) &= 8 \pi M_1(n) = 8 m_1 \pi n^{\lambda_1}\\
        s_2(n) &= 4 \pi M_0(n) = 4 \pi n
    \end{align}
\end{subequations}

At large $n$, \cref{eq:proba} fixes the value $m_3 = 3 m_1 m_2$
and we have the following set of closed equations for $\alpha \in \{1, 2\}$
\begin{subequations}
    \begin{align*}
        m_1 \lambda_1 &= \int_0^\infty \exp(-\frac{3 m_2 x + 3 m_1 x^2 + x^3}{3 m_1 m_2}) \dd{x} \\
        m_2 \lambda_2 &= 2 \int_0^\infty x \exp(-\frac{3 m_2 x + 3 m_1 x^2 + x^3}{3 m_1 m_2}) \dd{x}.
    \end{align*}
\end{subequations}
A root-finding algorithm finds
\begin{equation*}
    \lambda_1 \approx 0.6313, \quad \frac{m_2}{m_1^2} \approx 2.336.
\end{equation*}

\subsection{Dimension $d=4$}

In four dimensions, the surface function is given by
\begin{equation}
    S_n(r) = -4 V_4 \int_0^\infty (r+r')^3 \dv{\N{n}}{r'} \dd{r'}
\end{equation}
where $V_4$ is the volume of the four-dimensional unit sphere.
$\myset = \{0, 1, 2, 3\}$ with at large $n$
\begin{subequations}
    \begin{align}
        s_0(n) &= 4 V_4 M_3(n) = 4 V_4  m_3 n^{\lambda_3} \\
        s_1(n) &= 12 V_4 M_2(n) = 12 V_4 m_2 n^{\lambda_2} \\
        s_2(n) &= 12 V_4 M_1(n) = 12 V_4 m_1 n^{\lambda_1} \\
        s_3(n) &= 4 V_4 M_0(n) = 4 V_4  n.
    \end{align}
\end{subequations}
One obtains a closed set of equations for $\alpha \in \{1, 2, 3\}$
\begin{subequations}
    \begin{align*}
        m_1 \lambda_1 &= \int_0^\infty \exp(-\frac{4 m_3 x + 6 m_2 x^2 + 4 m_1 x^3 + x^4}{4 m_1 m_3 + 3 m_2^2}) \dd{x} \\ \\
        m_2 \lambda_2 &= 2 \int_0^\infty x \exp(-\frac{4 m_3 x + 6 m_2 x^2 + 4 m_1 x^3 + x^4}{4 m_1 m_3 + 3 m_2^2}) \dd{x} \\
        m_3 \lambda_3 &= 3 \int_0^\infty x^2 \exp(-\frac{4 m_3 x + 6 m_2 x^2 + 4 m_1 x^3 + x^4}{4 m_1 m_3 + 3 m_2^2}) \dd{x}.
    \end{align*}
\end{subequations}
A root-finding algorithm finds
\begin{equation*}
    \lambda_1 \approx 0.7369, \quad \frac{m_2}{m_1^2} \approx 1.6428, \quad \frac{m_3}{m_1^3} \approx 4.6867.
\end{equation*}

\section{Identical twins model in higher dimensions}
\label{app:twins}

\subsection{Dimension $d=3$}

Since it is an odd dimension, the surface function only contains integer powers of $r$.
The decomposition of $\surface{n}(r)$ is only modified for $\alpha = 1$ so that
\begin{subequations}
    \begin{align}
        s_0(n) &= 4 \pi M_2(n) = 4 m_2 \pi n^{\lambda_2}\\
        s_1(n) &= 6 \pi M_1(n) = 6 m_1 \pi n^{\lambda_1}\\
        s_2(n) &= 4 \pi M_0(n) = 4 \pi n
    \end{align}
\end{subequations}
The power-law parameters satisfy the following closed-set of equations for $\alpha \in \{1, 2, 3\}$
\begin{subequations}
    \begin{align*}
        m_1 \lambda_1 &= \int_0^\infty \exp(-\frac{3 m_2 x + 9 m_1 x^2 / 4 + x^3}{m_3}) \dd{x} \\
        m_2 \lambda_2 &= 2 \int_0^\infty x \exp(-\frac{3 m_2 x + 9 m_1 x^2 / 4 + x^3}{m_3}) \dd{x} \\
        -m_3 \lambda_3 &= 3 \int_0^\infty x^2 \exp(-\frac{3 m_2 x + 9 m_1 x^2 / 4 + x^3}{m_3}) \dd{x}.
    \end{align*}
\end{subequations}
A root-finding algorithm finds
\begin{equation*}
    \lambda_1 \approx 0.6285, \quad \frac{m_2}{m_1^2} \approx 2.4071, \quad \frac{m_3}{m_1^3} \approx 6.6972.
\end{equation*}
This is consistent with the constraint coming from \cref{eq:proba}
\begin{equation}
    m_3 = \frac{3 m_1 m_2 (10 \lambda_1 -3)}{4 (3 \lambda_1-1)}.
\end{equation}

\subsection{Dimension $d=4$}

In four dimensions, the identical twins model adds a half-integer power of $r$ to the decomposition of $\surface{n}(r)$
\begin{subequations}
    \begin{align}
        s_0(n) &= 4 V_4 M_3(n) = 4 V_4  m_3 n^{\lambda_3} \\
        s_1(n) &= 12 V_4 M_2(n) = 12 V_4 m_2 n^{\lambda_2} \\
        s_{3/2}(n) &= - \frac{16 \sqrt{2}}{3 \pi} V_4 M_{3/2}(n) = - \frac{16 \sqrt{2}}{3 \pi} V_4 m_{3/2} n^{\lambda_{3/2}} \\
        s_2(n) &= 12 V_4 M_1(n) = 12 V_4 m_1 n^{\lambda_1} \\
        s_3(n) &= 4 V_4 M_0(n) = 4 V_4  n.
    \end{align}
\end{subequations}
The power-law parameters satisfy the following closed-set of equations for $\alpha \in \{1, 3/2, 2, 3\}$
    \begin{subequations}
        \begin{align*}
            m_1 \lambda_1 &= \int_0^\infty \exp(-\frac{x^4 + 4 m_1 x^3 - \frac{32\sqrt{2}}{15\pi} m_{3/2} x^{5/2} + 6 m_2 x^2 + 4 m_3 x}{m_4}) \dd{x} \\
            m_{3/2} \lambda_{3/2} &= \frac{3}{2} \int_0^\infty \sqrt{x}  \exp(-\frac{x^4 + 4 m_1 x^3 - \frac{32\sqrt{2}}{15\pi} m_{3/2} x^{5/2} + 6 m_2 x^2 + 4 x m_3}{m_4}) \dd{x} \\
            m_2 \lambda_2 &= 2 \int_0^\infty x  \exp(-\frac{x^4 + 4 m_1 x^3 - \frac{32\sqrt{2}}{15\pi} m_{3/2} x^{5/2} + 6 m_2 x^2 + 4 m_3 x}{m_4}) \dd{x} \\
            m_3 \lambda_3 &= 3 \int_0^\infty x^2  \exp(-\frac{x^4 + 4 m_1 x^3 - \frac{32\sqrt{2}}{15\pi} m_{3/2} x^{5/2} + 6 m_2 x^2 + 4 m_3 x}{m_4}) \dd{x} \\
            - m_4 \lambda_4 &= 4 \int_0^\infty x^3  \exp(-\frac{x^4 + 4 m_1 x^3 - \frac{32\sqrt{2}}{15\pi} m_{3/2} x^{5/2} + 6 m_2 x^2 + 4 m_3 x}{m_4}) \dd{x}.
        \end{align*}
    \end{subequations}
    A root-finding algorithm finds
    \begin{equation*}
        \lambda_1 \approx 0.7362, \quad \frac{m_{3/2}}{m_1^{3/2}} \approx 1.2208, \quad \frac{m_2}{m_1^2} \approx 1.6611, \quad \frac{m_3}{m_1^3} \approx 4.8320 , \quad \frac{m_4}{m_1^4} \approx 26.522.
    \end{equation*}

\end{document}